\documentclass[reprint,superscriptaddress,amsmath,amssymb,aps]{revtex4-1}
\usepackage[utf8]{inputenc}
\usepackage[T1]{fontenc} 
\usepackage{graphicx}
\usepackage{rotating}
\usepackage{makeidx}
\usepackage{threeparttable}
\usepackage{float}
\usepackage{amsmath}
\usepackage{rotating}
\usepackage{hyperref}

\usepackage{relsize}
\usepackage{etoolbox}
\usepackage[most]{tcolorbox}
\usepackage{hyperref}
\usepackage{dcolumn}
\usepackage{bm}
\usepackage{wrapfig}
\usepackage{float}  
\usepackage[toc,page]{appendix}
\usepackage{gensymb}
\usepackage{epstopdf}
\usepackage{subfiles}
\usepackage{comment} 
\usepackage[autostyle=true]{csquotes} 
\usepackage{makecell}
\usepackage{siunitx}
\usepackage[all]{hypcap} 
\usepackage[most]{tcolorbox}
\usepackage{hyperref}
\usepackage{xcolor}
\usepackage{nicefrac}
\usepackage{chemformula}
\usepackage{ulem}  
\definecolor{gainsboro}{rgb}{0.86, 0.86, 0.86}
\hypersetup{
  colorlinks   = true, 
  urlcolor     = blue, 
  linkcolor    = blue, 
  citecolor   = blue 
}

\newcommand{\orbs}{  $ d_{x^2-y^2}, d_{3z^2-r^2} $  }

\begin{document}

\title{A non-perturbative study of bulk photovoltaic effect enhanced by an optically induced phase transition}

\author{Sangeeta Rajpurohit}
\email{srajpurohit@lbl.gov}
\affiliation{Molecular Foundry, Lawrence Berkeley National Laboratory, USA}

\author{C. Das Pemmaraju}
\affiliation{Stanford Institute for Materials \& Energy Sciences,
SLAC National Accelerator Laboratory, USA}

\author{Tadashi Ogitsu}
\affiliation{Lawrence Livermore National Laboratory, Livermore, USA}

\author{Liang Z Tan }
\affiliation{Molecular Foundry, Lawrence Berkeley National Laboratory, USA}

\date{\today}

\begin{abstract}
Solid systems with strong correlations and interactions under
light illumination have the potential for exhibiting interesting bulk
photovoltaic behavior in the non-perturbative
regime, which has remained largely unexplored in the past
theoretical studies. We investigate the bulk photovoltaic
response of a perovskite manganite with strongly coupled
electron-spin-lattice dynamics, using real-time
simulations performed with a tight-binding model. The transient
changes in the band structure and the photoinduced phase transitions,
emerging from spin and phonon dynamics, result in a nonlinear
current versus intensity behavior beyond the perturbative limit.
The current rises sharply across a photoinduced magnetic phase
transition, which later saturates at higher light intensities
due to excited phonon and spin modes. The predicted peak
photoresponsivity is orders of magnitude higher than other known
ferroelectric oxides such as BiFeO$_3$. We disentangle phonon-and
spin-assisted components to the ballistic photocurrent,
showing that they are comparable in magnitude. Our results
illustrate a promising alternative way for controlling and
optimizing the bulk photovoltaic response through the
photoinduced phase transitions in strongly-correlated systems. 
\end{abstract}
\maketitle

The bulk photovoltaic effect (BPVE) is the generation of photocurrent in
the bulk of a material, in the absence of any extrinsic carrier separation
mechanism such as heterojunctions, and arising purely from its intrinsic
noncentrosymmetry. As BPVE is not subject to the usual constraints of
conventional p-n junction photovoltaics, such as the Shockley-Queisser limit,
it has promising applications in next-generation light-harvesting and sensing.
Shift currents and phonon-induced ballistic currents are two main mechanisms
that have been put forward to explain BPVE in recent years, with relative
magnitudes that are strongly dependent on crystal structure and materials
system~\cite{Burger2019}. Shift currents~\cite{Baltz1981,Sipe2000,Young2012,Young2012_2,Tan2016}
arise from the asymmetry of the nonlinear interactions of carriers with
the light field. The ballistic currents are caused by the asymmetry of the
momentum distributions of charge carriers \cite{Belinicher1980,Sturman2020,Dai2021}
in non-centrosymmetric materials. 

Complex materials with tuneable interactions and correlations
are a promising class of systems to control and optimize
the BPVE. Previous theoretical studies of BPVE are
mostly based on perturbative methods
\cite{Young2012,Young2012_2,Cook2017,Gong2018,Dai2021}.
These theories fail to describe photovoltaic behavior of
such materials in non-perturbative regime where the charge
dynamics is strongly coupled to phonon and spin dynamics.
In this paper, we take a non-perturbative approach for studying the BPVE,
to investigate its behavior in a strongly-correlated system as light intensity
is increased continuously from the perturbative regime to current saturation.
While perturbative calculations of the shift and ballistic currents are mostly
performed by keeping the band structure and scattering matrix of carriers fixed,
this rigid-band approximation can potentially break down as the light intensity
is increased. For instance, electron-phonon interactions can result in deviations
from lowest-order behavior~\cite{Gong2018,Barik2020}. This is of particular
relevance for strongly-correlated systems with charge carriers dressed in
the form of excitons, polarons, and magnons
\cite{Kaindl2003,Lein2008, Buczek2009,Fischer2009,Verdi2017,Kampfrath2011,Reticcioli2019}.
To include the dynamical effects~\cite{Daranciang2012,Priyadarshi2012} arising due to
the changes in the local symmetries and the band structure, we simulate
the real-time evolution of the electron, spin and lattice degrees of freedom.

\begin{figure}[t]
     \begin{center}
     \includegraphics[width=\linewidth]{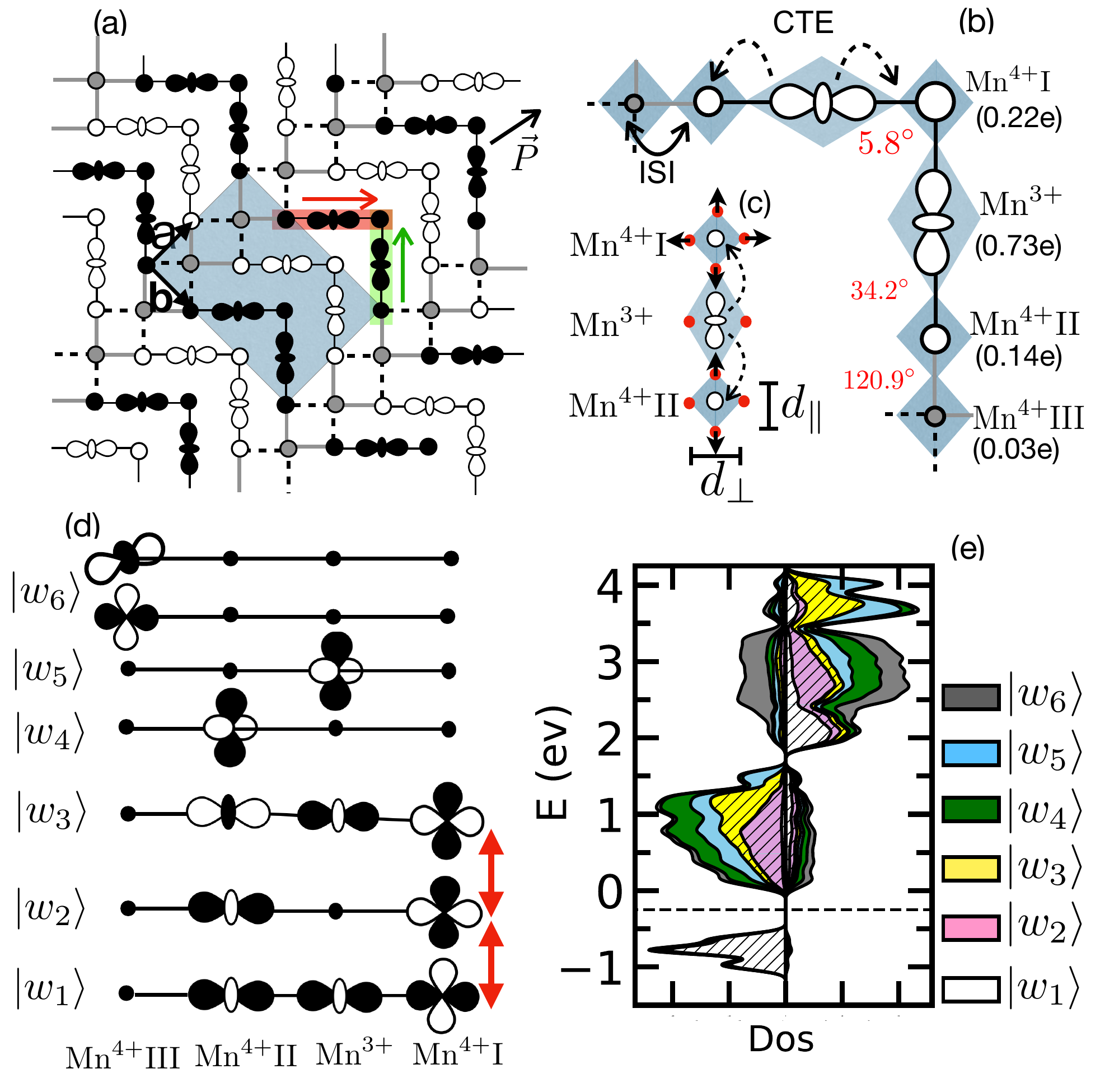}
     \end{center}
     \vspace{-0.5 cm}
     \caption{a): CO, OO, and SO of BS phase within the
    $ab$-plane. The circles and $d_{3x^2-r^2}/d_{3x^2-r^2}$-like
    orbitals are Mn$^{4+}$ and Mn$^{3+}$ sites, respectively.
    Black, white and gray indicates sites with different
    $t_{2g}$-spin orientation. The red (green) arrow indicates
    the local polarization direction of the horizontal (vertical)
    trimer and the black arrow indicate the net polarization
    $\vec{P}$ in the $ab$-plane. b): enlarged view of a pair of
    orthogonal adjacent trimer segments
    $\rm Mn^{4+}$ I -- $\rm Mn^{3+}$ -- $\rm Mn^{4+}$ II with
    additional $\rm Mn^{4+}$ III site indicating
    local e$_g$-electron densities (values in brackets) and
    intersite spin-angles (values in red).
    The solid and dashed arrows indicate charge transfer excitation (CTE)
    and photoinduced inter-site interaction (ISI).
    c): displacement (shown by arrows) of oxygen ions (red circles)
    following the photo-excitation. d): nature of Wannier states
    $|w_i\rangle$ where $i\in\{1,2,...6\}$. The red-arrows show
    inter-trimer dipole-allowed transitions. e): density of states
    for e$_g$ d-states of Mn atoms projected on $|w_i\rangle$-states.
    The right- and left-side shows majority- and minority-spin states,
    respectively and the dashed line is the Fermi level.}
    \label{fig:figures_BS1}
\end{figure}

We study the BPVE of a strongly-correlated perovskite manganite
$\rm{A_{1-x}A'_xMnO_3}$ (A=La, Pr and A'=Ca) at doping $x=2/3$
\cite{Mori1998,Radaelli1997,Fernandez1999}. In its ground state, this system
is in a  bi-stripe (BS) phase, and is an improper ferroelectric with
a weak polarization, exhibiting charge order (CO) and orbital order (OO). 
We report that its photocurrent has a strongly nonlinear dependence
on intensity, stemming from a photoinduced magnetic phase transition
that occurs during the optical excitation itself. Couplings to charge,
spin, and phonon degrees of freedom are each necessary for this
photoinduced phase transition. These results show that the
consideration of phonon and spin dynamics are necessary for designing
materials with tunable BPVE properties.

In A$_{1-x}$A'$_x$MnO$_3$ (A=rare-earth metal and A'=alkaline-earth metal)
the octahedral crystal field splits the Mn 3d-shell into three nonbonding
$t_{2g}$-orbitals and two antibonding e$_g$-orbitals. We treat the
delocalized e$_g$-orbitals using a tight-binding model, and the
localized $t_{2g}$-electrons as classical spins $\vec{S}_R$ with
length $|S|=3/2$ for each Mn-ion R.

The potential energy of the system is expressed as 
\begin{eqnarray}
E_{pot}\Big(|\psi_n\rangle,\vec{S}_{R},Q_{i,R}\Big)
&=&E_{e}(|\psi_n\rangle) +E_{S}(\vec{S}_R)+E_{ph}(Q_{i,R})\nonumber\\
&&\hspace{-2cm}+
E_{e-ph}(|\psi_n\rangle,Q_{i,R})+E_{e-S}(|\psi_n\rangle,\vec{S}_R)
 \label{eq:tbm}
\end{eqnarray} 
in terms of one-particle states 
$|\psi_n\rangle{=}\sum\limits_{\sigma,\alpha,i} |\chi_{\sigma,\alpha,i}\rangle \psi_{\sigma,\alpha,i,n}$
of e$_g$-electrons, $t_{2g}$-spin $\vec{S}_R$
and oxygen octahedral $Q_{i,R}$ phonon modes. The basis set
$|\chi_{\sigma,\alpha,i}\rangle$'s for the one-particle states
consists of local spin-orbitals with spin
$\sigma \in \{\uparrow,\downarrow\}$ and the orbital character $\alpha \in$ \orbs. 

The e$_g$-electrons delocalize between Mn sites via intermediate oxygen
bridges. The electrons experience an onsite Coulomb interaction,
a Hund’s coupling from $t_{2g}$-spin $\vec{S}_R$ and an electron-phonon
(el-ph) coupling with three local phonon modes per Mn-site.
The phonon modes are the octahedral breathing Q$_{1,R}$, and the two
Jahn-Teller (JT) active modes Q$_{2,R}$ and Q$_{3,R}$. These
modes are highly cooperative due to the oxygen atoms shared
between adjacent MnO$_6$ octahedra. The $t_{2g}$-spins $\vec{S}_R$
encounters an antiferromagnetic intersite coupling. For the complete
details of the model and its parameters, we refer to
\cite{Sotoudeh2017,Rajpurohit2020,Rajpurohit2020_2}.

We predict that the BS phase has a stable
non-collinear spin order (SO) in its ground state, illustrated
in top-left Figure\,\ref{fig:figures_BS1}, which is lower
in energy than previously reported collinear SOs \cite{Hotta2002_3}.
The ordering within $ab$-planes can be seen as an arrangement
of horizontal and vertical `trimers', which are three almost
ferromagnetically aligned Mn-sites, with average spin angles
of $5.8^{\circ}$ and $34.2^{\circ}$, in a row
$\rm Mn^{4+}$ I -- $\rm Mn^{3+}$ -- $\rm Mn^{4+}$ II. 
There are additional $\rm Mn^{4+}$ III sites that are not part of trimers. 
In the $\vec{c}$ direction, the SO is antiferromagnetic.
The BS ground state is noncentrosymmetric with a calculated
net polarization of 15\,nC/cm$^{2}$ in the $\vec{a}$ direction
(more information in SI).

The electronic structure of the BS phase can be
explained in terms of Wannier states
$|w_{i}\rangle$ ($i$ $\in\{1,2...6\}$) (details in SI),
spanning the Hilbert space of the trimer
segments and Mn$^{4+}$ sites, see Figure\,\ref{fig:figures_BS1}d
(more details in SI). The band-gap between the occupied $|w_1\rangle$-states
and the other unoccupied $|w_{i}\rangle$-states arises predominantly from the
JT splitting at Mn$^{3+}$ sites and is highly sensitive to the octahedral modes.   
The doubly degenerate $|w_{6}\rangle$ are two e$_g$-orbitals
of Mn$^{4+}$ III sites. The density of states
projected on the  $|w_{i}\rangle$-states, shown
in Figure\,\ref{fig:figures_BS1}e, clearly indicates
the valence band consists of the bonding state
of trimers, i.e., $|w_1\rangle$-states, with maximum
weight on the central $\rm Mn^{3+}$ sites. Above the
Fermi level are the nonbonding $|w_2\rangle$-states
localized on the $\rm Mn^{4+}$ I and II terminal sites.
Under optical excitation, the BS phase is expected to display 
two types of intra-trimer dipole-allowed transitions: 
$|w_1\rangle$-to-$|w_2\rangle$ and $|w_2\rangle$-to-$|w_3\rangle$.

We study the dynamics of the optically excited system by 
combining the model in Equation\,\ref{eq:tbm} with
Ehrenfest dynamics, as discussed in reference
\cite{Rajpurohit2020,Rajpurohit2020_2}. The effect of
the light-pulse, which is defined by electric field
$\vec{E}(r,t)=\vec{e}_A\omega \rm{Im}(A_oe^{-i\omega t})g(t)$,
is incorporated in the model by the Peierls
substitution method \cite{Peierls1933}. Here $A_o$
is the magnitude of the vector potential, $\omega$ is
the angular frequency, $\vec{e}_A$ is the direction of
the electric field. The pulse shape is Gaussian fixed by
$g(t)=e^{\frac{-t^2}{2c^2_{\omega}}}(\sqrt{\pi c^2_{\omega}})^{-1}$
where pulse duration defined by its FWHM$=2c_{\omega}\sqrt{\ln{2}}$.
The propagation of the single-particle wave functions
$|\psi_{\sigma,\alpha,R,n}\rangle$ and the spins
$\vec{S}_R$ is governed by time-dependent Schrodinger
equation while the atoms are treated classically
and they evolve according to Newton's equations of motion (more information in SI).
\begin{figure}[t]
     \begin{center}
     \includegraphics[width=\linewidth]{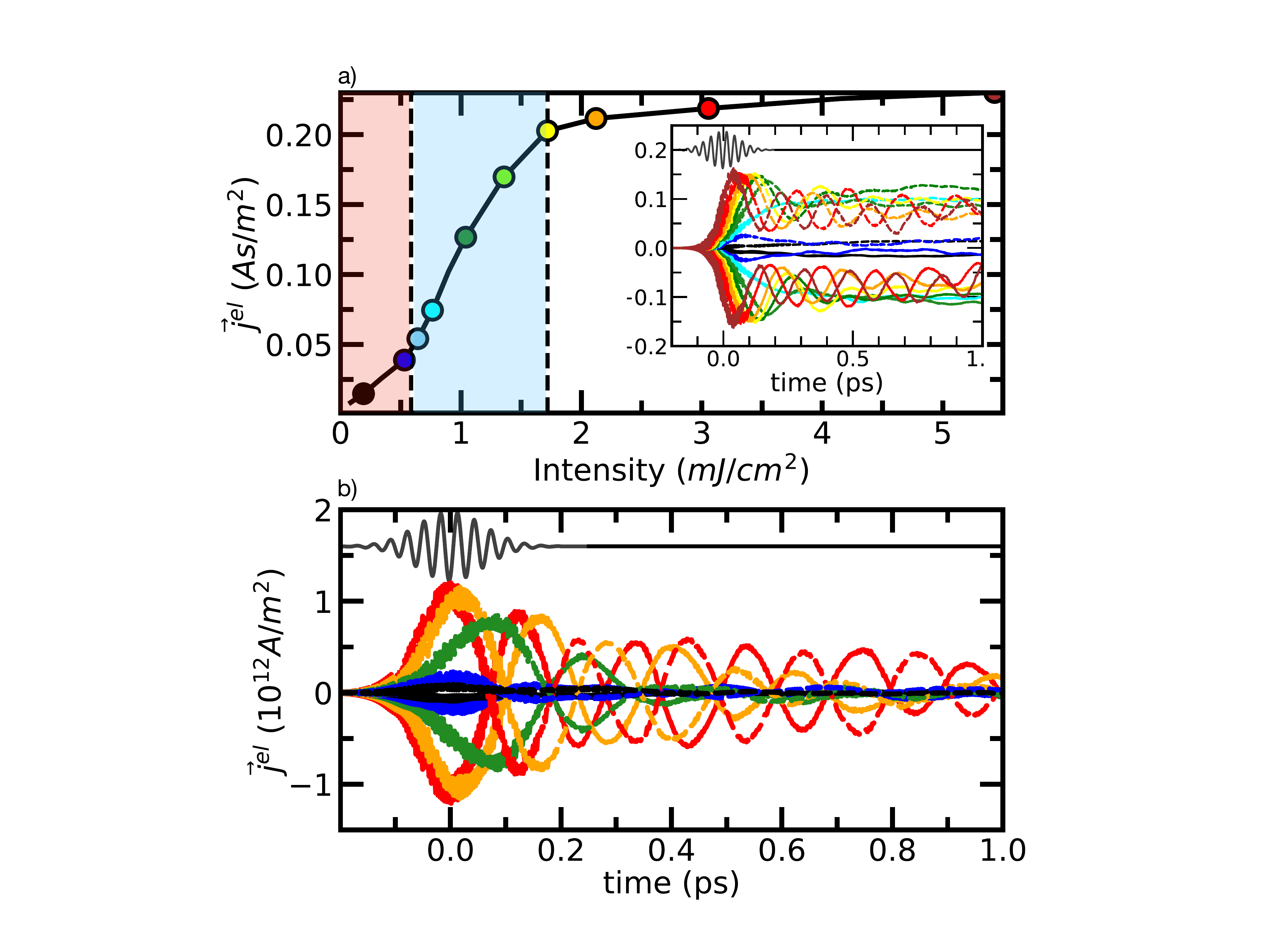}
     \caption{Generation and evolution of photocurrent a):
     Integrated current-density $\int dt j^{el} (t) $
     versus intensity $I$ showing three regions, namely I (red-shaded),
     II (blue-shaded), and III (white-shaded) with distinct photocurrent
     behavior for light polarized along $\vec{a}$ direction. The
     time-integration is performed from the beginning till the first
     $\vec{j}^{el}$ maximum. Inset shows the integrated current $\int dt j^{el} (t) $
     versus time $t$. The symbol and line colors refer to the different
     light intensities. b): Instantaneous current $j^{el} (t)$ versus time $t$. The
     light-pulse is shown in black in the top. The solid-and dashed-lines
     in the inset of left figure and in right figure show
     the direction $(\vec{a}+\vec{b})$ and $(\vec{a}-\vec{b})$,
     respectively, of the current. The colors refer to the
     increasing intensities with $A_o$=0.15\,$\hbar/ea_o$ (black),
     0.25\,$\hbar/ea_o$ (blue), 0.35\,$\hbar/ea_o$ (green),
     0.50\,$\hbar/ea_o$ (orange) and 0.60\,$\hbar/ea_o$ (red)
     The corresponding light-intensities are $I=0.19$
     mJ/cm$^2$ (black), 0.53 mJ/cm$^2$ (blue),
     1.04 mJ/cm$^2$ (green) and 2.12
     mJ/cm$^2$ (orange) and 3.05 mJ/cm$^2$ (red).}
     \label{fig:figures_BS2}
     \end{center}
\end{figure}

To measure the photocurrent magnitude, we calculate the 
evolution of the current-density $\vec{j}^{el}(t)$ which is
defined at time $t$ as
$\vec{j}^{el}(t){=}-\frac{e}{V}\sum\limits_{l\epsilon N_R}\vec{j}_l(t)$,
where $V{=}d_{Mn-Mn}^3N_R$ is the volume
of the unit cell with $N_R$ number of Mn ions and with
average bond length $d_{Mn-Mn}{=}\SI{3.84}{\angstrom}$ between
Mn ions. The current vector $\vec{j}_l(t)$ at lattice site $l$
is expressed as
\begin{eqnarray}
 \vec{j}_l(t)&=&\sum_n f_n(t)\sum\limits_{l' \epsilon \langle NN\rangle}  \sum_{\sigma}\sum_{\alpha,\beta}\frac{i}{\hbar}\Big(\psi^*_{\sigma,\alpha,l,n} (t) 
 T_{\alpha,\beta, l,l'}\psi_{\sigma,\beta,l',n}  (t) \nonumber \\
    &-&\psi^*_{\sigma,\beta,l',n} (t)T_{\beta,\alpha,l',l}\psi_{\sigma,\alpha,l,n} (t)  \Big) d_{Mn-Mn}\vec{e}_{l-l'}. 
    \label{eq:current}
\end{eqnarray}
Here, $ f_n(t)$ is the instantaneous occupation of the
one-particle states
$|\psi_n\rangle{=}\sum\limits_{\sigma,\alpha,l} |\chi_{\sigma,\alpha,l}\rangle \psi_{\sigma,\alpha,l,n} (t)$
and  $\vec{e}_{l-l'}=\Big(\frac{\vec{R_l}-\vec{R_{l'}}}{|\vec{R_l}-\vec{R_{l'}}|}\Big)$
is the unit vector in the direction joining sites $l$ and $l'$.
$T_{\alpha,\beta, l,l'}$ is the hopping matrix element between
e$_g$-orbitals $\alpha$ and $\beta$ at sites $l$ and $l'$, respectively. 

We simulate the photocurrent of a $12{\times}12{\times}4$
perovskite supercell under a 100\,fs Gaussian pulse and
periodic boundary conditions ($\Gamma$-point $k$-point sampling).
The photon energy was set at $\hbar\omega{=}0.97$ eV, where
the system shows maximum absorption. Here, we discuss results
for light polarization in the $\vec{a}$ direction, with light
polarization in the $\vec{b}$ direction being qualitatively
similar, resulting in photocurrents that are always aligned
along the bulk polarization direction. Light polarization
in the $\vec{c}$ direction results in no net current because
it contains a mirror plane. The current-vs-intensity behavior
over a range of intensities typical of ultrafast spectroscopies
separates into three distance regions ( Figure\,\ref{fig:figures_BS2}a):
region I ($I{=}0{-}0.65$ mJ/cm$^2$) where the current grows
linearly with the light intensity; region II ($I{=}0.65{-}1.70$ mJ/cm$^2$)
with a sharp rise in the photocurrent; region III
($I{=}1.70{-}$ mJ/cm$^2$ when the photocurrent saturates. 

In region I, lowest order BPVE is active; the linear dependence
of current on intensity shows that processes that are second order
in the electric field, namely shift and ballistic photocurrents,
contribute to the photocurrent. The current direction at very low
intensities is predominantly along the bulk polarization direction,
$\vec{a}$. Spin dynamics are not yet active in this region, with
the ground state SO remaining intact and the photoinduced dynamics
is entirely driven by electrons and phonons. The electronic
transitions in region I are predominantly  intra-trimer
$|w_1\rangle$-to-$|w_2\rangle$ and involve the charge transfer
from the $\rm Mn^{3+}$ to $\rm Mn^{4+}$ I and II sites of the
trimer, indicated by the solid arrow in  Figure\,\ref{fig:figures_BS1}b. 

\begin{figure}[t]
     \begin{center}
     \includegraphics[width=\linewidth]{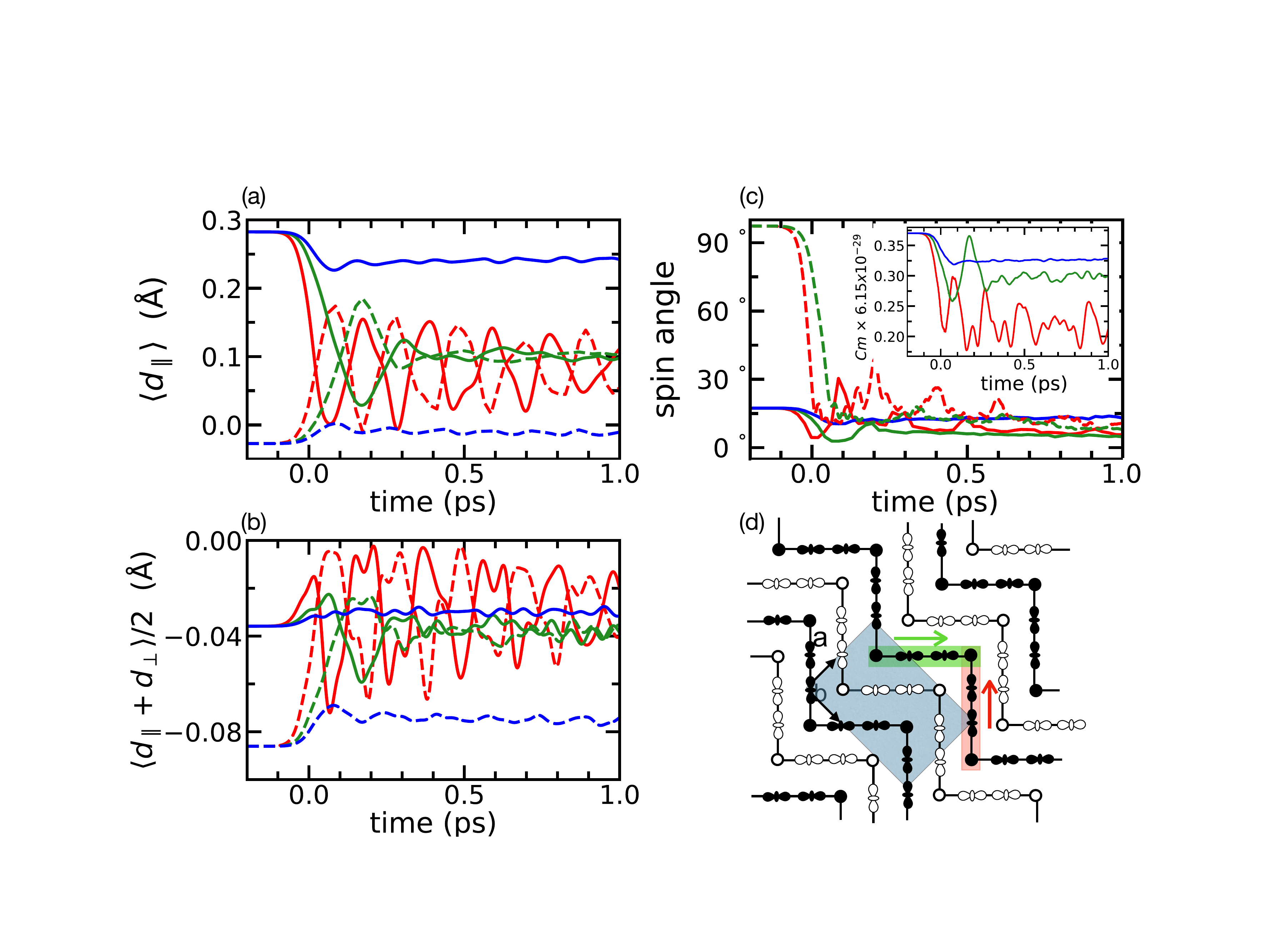}
     \caption{Dissipation of the excited-electrons energy into the
     octahedral phonon modes and t$_{2g}$-spins. a): O-O bond
     expansions $d_\parallel$ (see Figure \ref{fig:figures_BS1}c)
     along the local trimer axes around Mn$^{3+}$ (solid-lines),
     Mn$^{4+}$ I (dashed-lines). b): average O-O bond expansion
     $d_{\parallel_{ab}}=\langle\frac{d_\parallel+d_\perp}{2}\rangle$,
     around Mn$^{4+}$ II (solid-lines) and
     Mn$^{4+}$ III sites (dashed-lines). c):
     Evolution of the average intra-trimer spin angles (solid-lines)
     and inter-trimer spin angle along the direction of trimer
     (thick-lines). The inset shows evolution of average local dipole
     moment $P_l=\langle\sum\limits_{i}^{5}\rho_{i}\vec{r}_i\rangle$ of
     tetramer units, indicated in red and green in panel d.
     The colored lines refer to the increasing intensities with
     0.53 mJ/cm$^2$ (blue), 1.04 mJ/cm$^2$ (green) and 3.05 mJ/cm$^2$ (red). 
     d): Photo-induced collinear SO in the $ab$-plane.
     Spin-up (spin-down) sites are indicated by white (black).
     The $ab$ planes are antiferromagnetically coupled in $\vec{c}$ direction.}
     \label{fig:figures_BS3}
     \end{center}
\end{figure}

Oscillations in the instantaneous current over a 200 fs time
scale, which are driven by phonons, are already discernible in
region I. These oscillations do not contribute to the net
time-integrated current; to remove their contribution, we
plot only the time-integrated current $\int dt j^{el} (t)$
in the current-vs-intensity plot (Figure\,\ref{fig:figures_BS2}a). 
Due to the strong el-ph coupling, the octahedral phonon modes
follow the photoinduced alterations in the charge disproportion
between Mn sites. Oxygen octahedra around $\rm Mn^{3+}$ sites,
originally expanded in the ground state (Figure\,\ref{fig:figures_BS1}a),
start to shrink upon photoexcitation, as indicated by decreased
O-O bond distances $d_\parallel$ along the local trimer axes
in  Figure\,\ref{fig:figures_BS3}a. In contrast, the average
O-O bond-lengths in the $ab$-plane
around Mn$^{4+}$ II and III sites increase as shown in Figure\,
\ref{fig:figures_BS3}b. As soon as the
octahedral deformations around Mn sites adapt to the new local
charge densities, the coherent phonon modes of THz
frequencies are excited. The amplitude of the excited phonon
modes grows with the light intensity.

In the intensity region II (Figure\,\ref{fig:figures_BS2}a),
the current sharply increases due to a photoinduced magnetic
phase transition from the original non-collinear
SO to a new collinear SO, signaling a departure from the
regime of perturbative BPVE. In the new collinear SO,
$t_{2g}$ spins on Mn$^{4+}$ III sites rotate to bridge
spin-aligned trimers, originally disconnected in
the ground state, forming continuous ferromagnetic
zig-zag chains in the $ab$-plane, consisting of collinear horizontal
and vertical Mn-tetramers, as shown in Figure\,\ref{fig:figures_BS3}d.
The CO within tetramer lacks an inversion center causing asymmetric
intersite electron transfer between the original trimer and
the adjacent Mn$^{4+}$ III site. As a result of this new
ferromagnetic alignment in the non-centrosymmetric $ab$-plane,
carrier transport is enhanced, with the photocurrent being
predominantly in the $\vec{a}$ direction. 

The mechanism for the magnetic phase transition can be
understood as follows. The photo-induced intersite
electron-transfer between Mn$^{4+}$ II and III site via
$|w_1\rangle{-}|w_6\rangle$ and $|w_2\rangle{-}|w_6\rangle$
hybridization, along the trimer axes (shown by the solid black arrow
in Figure\,\ref{fig:figures_BS1}b) triggers
spin dynamics. This intersite electron-transfer is
restricted to the ferromagnetic spin component. 
This spin-restricted electron-transfer
changes the local e$_g$-electrons spin and perturbs
the $t_{2g}$-spins, due to Hund's coupling, inducing
the t$_{2g}$-spin dynamics. The evolution of the
$t_{2g}$-spin angles between sites
is shown in Figure\,\ref{fig:figures_BS3}c.
Above a critical intensity I$_o$=1.35 mJ/cm$^2$,
the perturbation on the $t_{2g}$-spins is strong enough
to induce a magnetic phase transition within 100 fs. 
The spin angle fluctuations settle within 500 fs. 

The electron-transfer within tetramers is sensitive to the
local orbital polarization and the onsite e$_g$-orbital energy
levels at Mn sites, which in turn depend on the phonon dynamics.
As long as there is a charge-disproportion between sites
within the tetramers, the phonon modes track the local charge
asymmetries over time, as shown in the inset of 
Figure\,\ref{fig:figures_BS3}c, and contribute to
phonon-induced photocurrent oscillations. These oscillations
decay on sub-picosecond timescales, primarily due to energy
loss to spin degrees of freedom.


Upon further increase in intensity (region III), the current starts 
saturating due to the dissipation of excited electron
energy into phonon modes. As seen in Figure~\ref{fig:figures_BS3}a-b,
phonon-induced current oscillations grow in amplitude in region III.
Further energy loss from the phonon modes into spin degrees of
freedom is also evident from the sub-picosecond decay of phonon
oscillations in Figure~\ref{fig:figures_BS3}a-b. A secondary effect
is the change in the electronic structure resulting from excited
phonon modes. Upon photoexcitation in region III, the energy gap
between the valence and conduction bands is reduced. This detunes
the resonance between the light field and the peak absorption energy,
which reduces the absorption coefficient and photocurrent at high intensities. 

Overall, the magnetic phase transition results in an enhancement
of photocurrent compared to other materials that display BPVE in
the perturbative limit. The photoresponsivity $R_p=\frac{1}{I}\int\vec{j}^{el}(t)dt$,
where the time-integration is performed over light-pulse
duration, calculated for intensity I=2.12 mJ/cm$^2$ at the time
of current saturation in region III is
$10$ mA/W, an order of magnitude higher the 
photoresponsivity for the known ferroelectric oxide BiFeO$_3$
\cite{Young2012_2}.


\begin{figure}[t]
     \begin{center}
     \includegraphics[width=\linewidth]{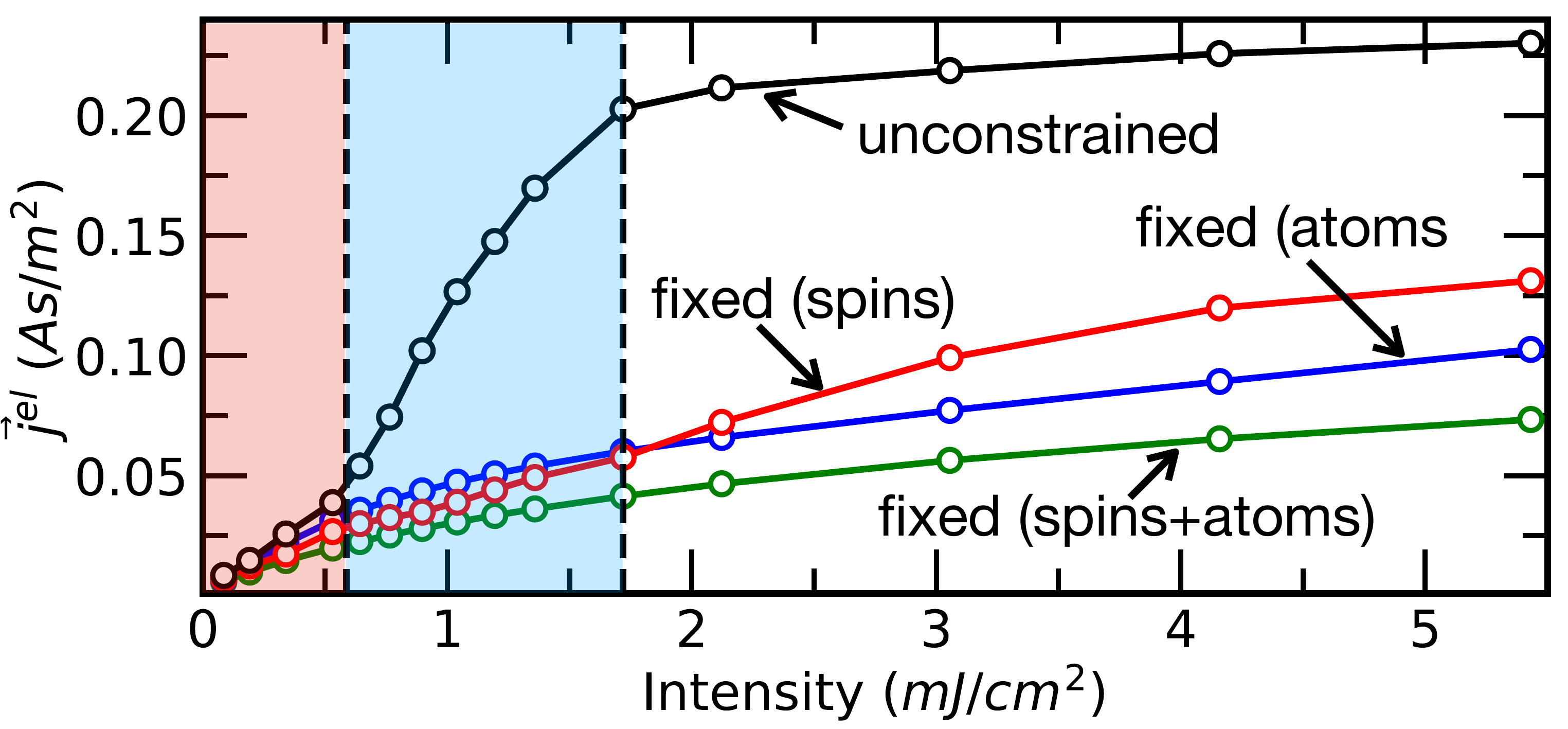}
     \caption{Integrated current-density $j^{el} (t)$-vs-intensity
     during the light-pulse for the different cases. The unconstrained
     case shows three intensity  regions, I (red-shaded), II (red-shaded),
     and III (white-shaded) with distinct current behavior.}
     \label{fig:figures_BS4}
     \end{center}
\end{figure}
To measure the contribution of the phonon and spin-assisted
processes, we repeat the simulations keeping either
atoms or spins or both fixed (Figure\,\ref{fig:figures_BS4}). 
With constrained spins, there is clearly no photoinduced magnetic
phase transition. Moreover, with fixed atoms, we do not find any
magnetic phase transitions either, for the light intensity range
covered in the present work, suggesting that phonon dynamics is
crucial for the phase transition. This is because the photoinduced
octahedral deformations bring $|w_1\rangle$, $|w_2\rangle$ and
$|w_6\rangle$ states, involved in the $|w_1\rangle{-}|w_6\rangle$ and
$|w_2\rangle{-}|w_6\rangle$ hybridization, energetically
closer, strengthening the intersite coupling between
$\rm Mn^{4+}$ II and III sites (see solid arrow in top-right
Figure\,\ref{fig:figures_BS1}), thus facilitating the
magnetic phase transition.  
The photocurrent in the cases with constrained atoms or spins is
linear with intensity, with no current saturation, implying that
they remain in the perturbative BPVE limit over the range of
simulated intensities.

In the absence of the spin and lattice dynamics, as in the fixed
(atoms+spins) case in Figure\,\ref{fig:figures_BS4}, the
photocurrent is purely the shift current along with other
higher-order contributions. The additional current in the fixed
(spins) and fixed (atoms) cases reflects the individual contribution
from the phonon-assisted and spin-assisted processes, respectively.
We see that both phonons and spins act to increase the overall
photocurrent significantly, even in this perturbative limit, with
the spin-assisted processes doubling the current of the fully
constrained case. At low light intensities (region I), the spin
contribution to the current exceeds the phonon contribution. However,
for higher intensities, the phonon-assisted ballistic current
becomes larger compared to its spin counterpart.
In the absence of phonon dynamics, as for the fixed (atom) case,
the excited electrons dissipate energy directly into
the spins generating the spin-assisted ballistic current,
see Figure\,\ref{fig:figures_BS4}. 
In all of these cases, the direction of current during the
light pulse remains predominantly along the bulk polarization direction.

Our results highlight the significance of the photoinduced
effects driven by spin and lattice dynamics in BPVE, suggesting
the potential for controlling photovoltaic materials responses
with tunable excitations and interactions. A design strategy
for large BPVE suggested by these results is to look for systems
with strong correlations that can undergo optically induced phase
transitions at low intensities producing an abrupt change in the
photocurrent like in the case studied here.
Compared to other nonpolar modes, the electronic
properties in complex oxides are strongly linked to
the JT distortions. Our study reveals that the coherent
dynamics of the excited JT and breathing modes contribute 
to the phonon-assisted ballistic photocurrent currents. Transition
metal oxides such as manganites or nicklates, with spin orders in
ground and excited states, are candidates to show spin-assisted
BPVE as demonstrated here. 

In conclusion, we have shown how a combination of spin- and
phonon-induced processes can substantially enhance the bulk
photovoltaic effect, using a non-perturbative methodology. 
The real-time simulations of a strongly correlated system show
that photoinduced phase transitions, which are generally ignored
in perturbative theoretical methods, significantly impact the
photocurrent generation and its evolution.
Understanding the effect of the transient spin
and lattice dynamics on the dynamical nature of
the band structure can be exploited for desirable
photovoltaic properties by tuning the correlations
and interactions in correlated systems through
targeted materials design.

S.R. was supported by the Computational
Materials Sciences Program funded by the US Department of Energy,
Office of Science, Basic Energy Sciences, Materials Sciences and
Engineering Division. L.Z.T. was supported by the Molecular
Foundry, a DOE Office of Science User Facility supported by the
Office of Science of the U.S. Department of Energy under Contract
No. DE-AC02-05CH11231.
\bibliography{ref}

\newpage
\end{document}